\documentclass[twocolumn,showpacs,preprintnumbers,amsmath,amssymb]{revtex4}

\usepackage{graphicx}
\usepackage{dcolumn}
\usepackage{bm}

\begin{document}

\title{Numerical analysis of quasinormal modes in nearly extremal Schwarzschild-de 
Sitter spacetimes}

\author{Shijun Yoshida}
\email{yoshida@fisica.ist.utl.pt}
\affiliation{
Centro Multidisciplinar de Astrof\'{\i}sica - CENTRA, 
Departamento de F\'{\i}sica, Instituto Superior T\'ecnico,
Av. Rovisco Pais 1, 1049-001 Lisboa, Portugal}

\author{Toshifumi Futamase}
\email{tof@astr.tohoku.ac.jp}
\affiliation{Astronomical Institute, Graduate School of Science,
           Tohoku University, Sendai 980-8578, Japan}

\date{\today}

\begin{abstract}

We calculate high-order quasinormal modes with large imaginary frequencies 
for electromagnetic and gravitational perturbations in nearly extremal 
Schwarzschild-de Sitter spacetimes. Our results show that for low-order quasinormal 
modes, the analytical approximation formula in the extremal limit derived by Cardoso 
and Lemos is a quite good approximation for the quasinormal frequencies as long as the 
model parameter $r_1\kappa_1$ is small enough, where $r_1$ and $\kappa_1$ are the 
black hole horizon radius and the surface gravity, respectively. For high-order 
quasinormal modes, to which corresponds quasinormal frequencies with large imaginary 
parts, on the other hand, this formula becomes inaccurate even for small 
values of $r_1\kappa_1$.  We also find that the real parts of the quasinormal 
frequencies have oscillating behaviors in the limit of highly damped modes, which are 
similar to those observed in the case of a Reissner-Nordstr{\" o}m black hole. 
The amplitude of oscillating ${\rm Re(\omega)}$ as a function of
${\rm Im}(\omega)$ approaches a non-zero constant
value for gravitational perturbations and zero for electromagnetic
perturbations in the limit of highly damped modes, where $\omega$ denotes the 
quasinormal frequency. This means that for gravitational perturbations, the 
real part of quasinormal modes of the nearly extremal Schwarzschild-de Sitter 
spacetime appears not to approach any constant value in the limit of highly damped 
modes.  On the other hand, for electromagnetic perturbations, the real part of 
frequency seems to go to zero in the limit. 

\end{abstract}

\pacs{04.70.Bw, 04.70.-s, 04.30.-w}

\maketitle

\section{Introduction}
The quasinormal modes (QNMs) of spacetimes containing black holes 
have been studied since the pioneering work of Vishveshwara \cite{vish}, who 
first observed quasinormal ringing of a Schwarzschild spacetime in his 
numerical calculations. 
The main motivation to study the QNM of a black hole is twofold: 
One is to answer the question of whether the spacetime is stable or not, and the other 
to know what kind of oscillations will be excited in the spacetime when  
some perturbations are given. In an astrophysical point of view, the latter is 
quite important from the observational point of view because we could 
determine fundamental parameters of a black hole, such as the mass or 
the angular momentum, through the information of the QNMs. Thus, a large number 
of studies on the QNMs of spacetimes containing black holes 
have been done (for review, see, e.g., \cite{kokkotas99,nollert99}).

There is another interesting aspect of the QNMs in a black hole which 
is related to the quantum theory of gravity. Bekenstein and Mukhanov 
discussed the relationship between the fundamental area unit in the quantum 
theory of gravity and a Bohr transition frequency, applying Bohr's 
correspondence principle with a hydrogen atom to the quantum theory of 
a black hole \cite{MB95,B97}. For a Schwarzschild black hole, they then 
derived the Bohr transition frequency $\omega$, given by
\begin{equation}
\omega = \ln k / 8\pi M \,,
\label{bekenstein} 
\end{equation}
where $k=2, 3, 4\cdots$, and $M$ stands for the mass of a black hole. 
Furthermore they predicted the value of $k$ to be 
$k=2$, and suggested that this frequency should be equal to classical 
oscillation frequencies of the black hole. 
A few years ago, Hod however noticed that if $k=3$, the frequency 
given by formula (\ref{bekenstein}) is in quite good agreement with the 
asymptotic frequency of the QNM of a Schwarzschild black hole in the 
limit of highly damped modes, and proposed to apply Bohr's correspondence 
principle in order to determine the value of the fundamental area unit 
in the quantum theory of gravity, namely the value of $k$ \cite{hod} (see, also 
\cite{dreyer}). Since Hod's
proposal, the QNMs with large imaginary frequencies of spacetimes
including black holes have been attracted much attention, and a lot of
papers related to this subject have appeared in order to see
whether Hod's conjecture is applicable not only for a Schwarzschild
black hole but also for other black hole spacetimes. For example, Motl
analytically obtained an asymptotic constant value of the QNM
frequencies of a Schwarzschild black hole \cite{motl}, which had
been obtained numerically by Nollert \cite{nollert}. Motl and
Neitzke \cite{motl03}, Berti and Kokkotas \cite{berti}, Neitzke \cite{an03}, 
and Andersson and Howls
\cite{andersson_howls} studied the asymptotic behaviors of the QNMs in a
Reissner-Nordstr{\" o}m black hole. Berti et al. \cite{bertietal}, and Hod
\cite{hod03} discussed the QNMs of a Kerr black hole in the limit of highly damped
modes. 
As for a Schwarzschild-de Sitter (SdS) black hole, Cardoso and Lemos
\cite{cardoso03} and Maassen van den Brink \cite{brink} analytically obtained
asymptotic form of the QNM frequencies in almost the extremal limit, in which
the cosmological horizon becomes very close to the black hole horizon 
(for the case of a Schwarzschild black hole in an anti-de Sitter spacetime, 
see, e.g., \cite{CKL03,K02}).

A large number of papers related to Hod's conjecture, which have recently 
appeared, suggest that Hod's conjecture is not universal, at least as it 
stands, even though it is applicable for Schwarzschild black holes in four 
and higher dimensions (for the higher dimensional case, see, e.g.,  
\cite{motl03,db03,an03,cly03}). For example, a 
real part of the quasinormal frequencies in a Reissner-Nordstr{\" o}m black hole 
appears not to go to any constant value in the limit of highly damped mode, 
but shows some periodic behaviors as the imaginary part is increased 
\cite{motl03,berti,an03,andersson_howls}. This means that Hod's conjecture is not 
applicable in the Reissner-Nordstr{\" o}m case because a real part of the 
quasinormal frequencies does not have a limit as the imaginary part goes to infinity. 
In such a situation, it is necessary to explore a problem whether or not there 
is another black hole spacetime in which Hod's conjecture is applicable. 
The purpose of this paper is to improve our understanding of this problem 
and we are concerned here with the QNMs of SdS spacetimes.  
SdS spacetime has no spatial infinity but has cosmological horizon, and if 
Hod's conjecture is applicable, it is interesting to see whether Hod's conjecture 
depends only on the black hole horizon, but not the cosmological horizon.
Furthermore, recent observations show that the universe does have a non-zero positive 
cosmological constant. Therefore, SdS spacetimes are considered to be a good simple 
model of a black hole in the universe.

In this study, in particular, we calculate numerically the QNMs of nearly 
extremal SdS spacetimes for reason we describe below. 
We therefore assume the surface gravity $\kappa_1$ at the 
black hole horizon to be $\kappa_1\le 10^{-2}\ r_1^{-1}$,
where $r_1$ stands for the coordinate radius of the black hole horizon. Note that 
the extremal limit of the SdS space corresponds to the limit of $\kappa_1\rightarrow 0$. 
In a nearly extremal SdS black hole, as mentioned before, an analytical formula for 
quasinormal frequencies can be derived \cite{cardoso03,brink}. One of the aims of this 
paper is to examine whether this analytical formula is correct in the limit of highly 
damped modes. 

The paper is organized as follows. 
In \S 2 we present the basic equations for obtaining QNMs 
in the SdS spacetime using Leaver's continued fraction technique \cite{leaver}, 
which was extended to the case 
of the SdS spacetime by Moss and Norman \cite{moss}.  
Numerical results are given in \S 3, and \S 4 is devoted for 
conclusion.

\section{Method of Solutions}
In order to examine the QNMs of the SdS spacetime, we make use of the same 
formalism as that derived by Moss 
and Norman \cite{moss}, who obtained low-order quasinormal frequencies of the SdS 
spacetime for a wide range of the model parameter. The line element of the SdS 
spacetime is given by
\begin{equation}
ds^2=-{\Delta\over r^2}\,dt^2+{r^2\over\Delta}\,dr^2+
r^2(d\theta^2+\sin^2\theta\,d\phi^2)\,, 
\label{metric}
\end{equation}
where 
\begin{equation}
\Delta=r^2-2Mr-{1\over 3}\Lambda r^4\,. 
\label{def-delta}
\end{equation}
Here, $M$ and $\Lambda$ stand for the mass of the black hole and the cosmological 
constant, respectively. If a non-negative cosmological constant is assumed, 
namely de Sitter spacetime, there are two horizons, whose radial coordinates are 
given as positive solutions of $r^{-1}\Delta=r-2M-{1\over 3}\Lambda r^3=0$, 
one is the black hole horizon $r=r_1$ and the other the cosmological horizon 
$r=r_2$, where $r_2>r_1>0$. Note that the equation $r-2M-{1\over 3}\Lambda r^3=0$ 
has one negative solution $r=r_3<0$ for the SdS space. One of the important model 
parameters of the SdS spacetime is the surface gravity $\kappa_1$, defined by
\begin{equation}
\kappa_1=\lim_{r\rightarrow r_1}{1\over 2r^2}{d\Delta\over dr}\,. 
\label{def-kappa}
\end{equation}
In terms of the non-dimensional surface gravity $r_1\kappa_1$, the mass and the 
cosmological constant can be written as 
\begin{equation}
M={1\over 3}r_1(r_1\kappa_1+1)\,,\quad \Lambda=r_1^{-2}(1-2r_1\kappa_1)\,, 
\end{equation}
which shows that $0<r_1\kappa_1<1/2$ for the SdS spacetime. In this study, 
we employ the non-dimensional parameter $r_1\kappa_1$ to specify the SdS 
spacetime.

By virtue of the symmetry properties of the SdS spacetime, the master equations 
for the scalar ($s=0$), electromagnetic ($s=1$) and gravitational perturbations ($s=2$) 
can be cast into a wave equation of the simple form, given by \cite{mm90,of91,cl01}
\begin{equation}
{d^2\phi(r)\over dr_*^2}+[\omega^2-V(r)]\phi(r)=0\,,
\label{wave-equation}
\end{equation}
where $r_*$ denotes the tortoise coordinate, defined by $dr_*=r^2\Delta^{-1}dr$, 
and $\omega$ is the oscillation frequency of the perturbations.  Depending on the 
type of perturbations, here, the effective potential is explicitly given by  
\begin{equation}
{r^4V\over\Delta}=\left\{
\begin{array}{ll}
l(l+1)+{2M\over r}-{2\Lambda r^2\over 3}&{\rm for\ } s=0 \\ 
l(l+1)&{\rm for\ } s=1 \\
l(l+1)-{6M\over r}&{\rm for\ } s=2\,, 
\end{array}
\right.
\label{def-potential}
\end{equation}
where $l$ means the angular quantum number of perturbations. Here, only the 
axial parity perturbations have been considered for the gravitational case because 
quasinormal frequencies of the polar parity perturbations are the same as those 
of the axial parity perturbations (for the proof, see Appendix).

The QNMs of the SdS spacetime are characterized by the boundary 
conditions of incoming waves at the black hole horizon and outgoing waves at 
the cosmological horizon, given by 
\begin{equation}
\phi(r)\rightarrow\left\{
\begin{array}{ll}
e^{-i\omega r_*}&{\rm as}\ r_*\rightarrow \infty\\ 
e^{ i\omega r_*}&{\rm as}\ r_*\rightarrow-\infty\,,
\end{array}
\right.
\label{def-bc}
\end{equation}
where the time dependence of perturbations has been assumed to be $e^{i\omega t}$. 
In general, it is impossible to adapt the boundary condition (\ref{def-bc}) in 
a straightforward numerical integration to obtain quasinormal frequencies. Some 
special technique is therefore required for computations of QNMs. 
In the present investigation, we employ a standard technique devised by Leaver, namely 
the continued fraction method \cite{moss,leaver}.

To apply the continued fraction method to the SdS spacetime, it is convenient to 
introduce a new independent variable, defined by $x=r^{-1}$. With this new variable 
$x$, the asymptotic form of the perturbations as $r_*\rightarrow\infty$ can be 
rewritten as
\begin{equation}
e^{-i\omega r_*}=(x-x_1)^{-\rho_1}(x-x_2)^{-\rho_2}(x-x_3)^{-\rho_3}\,,
\label{expwr}
\end{equation}
where $x_i=r_i^{-1}$ and $\rho_i=i\omega/(2\kappa_i)$ for $i=1,2,3$, where 
\begin{eqnarray}
\kappa_1&=&M(x_1-x_2)(x_1-x_3)\,,\nonumber \\
\kappa_2&=&M(x_2-x_1)(x_2-x_3)\,, \\
\kappa_3&=&M(x_3-x_1)(x_3-x_2)\,.\nonumber 
\end{eqnarray}
The perturbation function $\phi$ is expanded around the black hole horizon as 
\begin{equation}
\phi=(x-x_1)^{\rho_1}(x-x_2)^{-\rho_2}(x-x_3)^{-\rho_3}\sum_{n=0}^\infty 
a_n\left({x-x_1\over x_2-x_1}\right)^n\,,
\label{expansion}
\end{equation}
where $a_0=1$ and $a_n$'s for $n\ge 1$ are determined by the three term recurrence 
relation, given by 
\begin{equation}
\alpha_na_{n+1}+\beta_na_n+\gamma_na_{n-1}=0\,,
\end{equation}
where 
\begin{eqnarray}
\alpha_n&=&2M(x_1-x_3)\{n^2+2(\rho_1+1)n+2\rho_1+1\}\,,\\
\beta_n&=&-2M(2x_1-x_2-x_3)\times\nonumber\\ 
&&\{n^2+(4\rho_1+1)n+4\rho_1^2+2\rho_1\}-l(l+1)\nonumber\\
&&+2Mx_1(s^2-1)\,,\\
\gamma_n&=&2M(x_1-x_2)(n^2+4\rho_1n+4\rho_1^2-s^2)\,. 
\end{eqnarray}
Here, $s=2$ for the gravitational perturbations and $s=1$ for the electromagnetic 
perturbations. Note that in the present study, we do not consider the scalar 
perturbations because Leaver's method cannot be directly applied to the scalar case. 
Comparing the expanded eigenfunction (\ref{expansion}) with equation (\ref{expwr}), 
we can see that the eigenfunction (\ref{expansion}) satisfies the QNM  
boundary condition (\ref{def-bc}) if the power series in equation (\ref{expansion}) 
converges for $x_2\le x\le x_1$. This convergence condition is equivalent to the 
condition written in terms of continued fractions \cite{gautschi}, which is given by 
\begin{eqnarray}
0=\beta_0-{\alpha_0\gamma_1\over\beta_1-}{\alpha_1\gamma_2\over\beta_2-}\cdots
\label{a-eq1}
\end{eqnarray}
Therefore, we solve this algebraic equation to obtain the quasinormal frequency.

\section{Numerical Results}
In this study, we are concerned with asymptotic behaviors of high-order QNMs, 
namely quasinormal frequencies with large imaginary parts, in almost the  
extremal limit, in which two horizons are quite near. We therefore consider only the 
case of $r_1\kappa_1\le 10^{-2}$. For these small values of $r_1\kappa_1$, the continued 
fractions (\ref{a-eq1}) converge very quickly and Leaver's method works quite well even 
when quasinormal frequencies have quite large imaginary parts. 
Note that for moderate values of $r_1\kappa_1$, however, the convergence of the continued 
fractions gets worse and Leaver's method is applicable only for modes with smaller imaginary 
frequencies. For those cases, thus, some other techniques such as Nollert's method \cite{nollert} 
or a phase integral method \cite{andersson} should be used to obtain high-order QNMs.

In order to check our numerical code, we have calculated fundamental frequencies of 
the QNMs for several values of $r_1\kappa_1$ and have fitted the mode 
frequencies as a function of $r_1\kappa_1$ with the polynomials defined by 
\begin{eqnarray}
{\rm Re}(\omega r_1)&=&r_1\kappa_1 b_0(1-b_1r_1\kappa_1)\,,\nonumber \\
{\rm Im}(\omega r_1)&=&r_1\kappa_1 c_0(1-c_1r_1\kappa_1)\,. 
\label{w-exp}
\end{eqnarray}
Recently, Cardoso and Lemos \cite{cardoso03} and Maasse van den Brink \cite{brink} 
analytically obtained the expansion coefficients, which are given by
\begin{equation}
b_0=\left\{
\begin{array}{ll}
\sqrt{l(l+1)-{1\over 4}}&{\rm for\ } s=0,\ 1\\
\sqrt{l(l+1)-{9\over 4}}&{\rm for\ } s=2 \,,
\end{array}
\right. 
\label{ana-coef1}
\end{equation} 
\begin{equation}
c_0=n+{1\over 2}\,,\quad b_1=c_1={2\over 3}\,,
\label{ana-coef2}
\end{equation}
where $n$ is the mode number. It is found that numerically obtained coefficients are 
in good agreement with coefficients given by (\ref{ana-coef1}) and (\ref{ana-coef2}). 
Note that our numerical results for the mode frequencies are consistent with those 
obtained by Moss and Norman \cite{moss}, who studied low-order QNMs 
of the gravitational perturbations for full range of the parameter $r_1\kappa_1$.

First, let us discuss properties of the quasinormal frequencies for the low-order 
modes. In Figure 1, we show the 
real parts of the frequencies for the low-order QNMs versus the imaginary 
parts of the frequencies for the gravitational perturbations with $l=2$. In this figure, 
non-dimensional frequencies ${\rm Re}(\omega/\kappa_1)$ have been plotted as a function of 
${\rm Im}(\omega/\kappa_1)$ for $r_1\kappa_1=10^{-3}$ and $r_1\kappa_1=5\times 10^{-3}$,  
and the dashed curve indicates the approximate frequency for $r_1\kappa_1\rightarrow 0$, 
derived by Cardoso and Lemos \cite{cardoso03} (see, also \cite{brink}). For low-order modes,  
Figure 1 illustrates how the frequencies of the QNMs behaves when the mode number and/or 
the value of $r_1\kappa_1$ is altered. Basic properties of the low-order QNMs 
are summarized as follows: For the modes associated with a small mode number, the real parts of 
the frequencies decrease with the increase of the imaginary parts of the frequency, even though 
the real parts of the frequency are constant in the approximation formula 
(\ref{w-exp})-(\ref{ana-coef2}). In other words, the 
analytical approximation formula in the extremal limit (\ref{w-exp})-(\ref{ana-coef2}) is quite 
good for the fundamental modes as long as $r_1\kappa_1$ is small enough, while, as expected in 
\cite{brink}, this approximation formula gets worse as the mode number is increased even for 
small values of $r_1\kappa_1$. This means that formula 
(\ref{w-exp})-(\ref{ana-coef2}) 
does not give a correct asymptotic value of the QNM frequencies in the limit of large imaginary 
frequencies. Similar properties can be seen for other perturbations 
having different $s$ and $l$. In Figure 2, we show the same results as those in 
Figure 1 but for the electromagnetic perturbations having $l=1$. It is observed  
that similar behaviors of the QNM frequencies are seen in the case of electromagnetic 
perturbations, too.

Now, we explain our numerical results for the asymptotic behavior of the QNM 
of nearly extremal SdS spacetimes in the limit of large imaginary 
frequencies. 
In Figure 3, we plot the imaginary parts of the non-dimensional 
QNM frequencies, $\omega/\kappa_1$, of the gravitational perturbations 
with $l=2$ as a function of the mode number $n$. 
In this figure, the model parameter is $r_1\kappa_1=10^{-3}$. 
The figure shows that an asymptotic form of ${\rm Im}(\omega/\kappa_1)\sim
n$, 
which is similar to the analytical 
formula (\ref{w-exp})-(\ref{ana-coef2}), is a good approximation for the imaginary 
parts in the limit of large imaginary frequencies. 
The same asymptotic form is inferred in all other QNM's 
we have calculated in the present study, 
regardless of the values of $r_1\kappa_1$, $l$, and $s$.

Let us next focus on the behaviors of the real part of the QNM
frequencies in the limit of highly damped modes. 
In Figures 4 and 5, the real parts of the non-dimensional mode frequencies, 
$\omega/\kappa_1$ are plotted 
as a function of the imaginary parts of the frequencies up to
sufficiently high-order modes for the $l=2,\ 3$ gravitational and 
for the $l=1,\ 2$ electromagnetic perturbations, respectively. 
The results for the model parameter of $r_1\kappa_1=10^{-3}$ are shown
in both figures. It is found that the real parts of the frequencies show 
oscillating behaviors as the imaginary parts of the frequencies are increased. 
It is important to note that similar oscillating behaviors have been 
observed in the QNMs of a Reissner-Nordstr{\" o}m black hole. (The quasinormal 
frequencies with large imaginary frequency of a Reissner-Nordstr{\" o}m black 
hole can be given in terms of a solution of the algebraic equation 
\cite{motl03,an03,andersson_howls}, 
\begin{equation}
e^{\beta\omega}+2+3e^{k^2\beta\omega}=0\,,
\label{rne}
\end{equation}
where $\beta$ and $k$ are contents determined with the mass and charge of the 
black hole (see, \cite{motl03}). As shown by Neitzke \cite{an03} and Anderson 
and Howls \cite{andersson_howls}, equation (\ref{rne}) has an infinite number 
of solutions and some solutions of equation (\ref{rne}) show periodicity.) 
The behavior of the amplitude of the oscillating 
${\rm Re}(\omega/\kappa_1)$ as a function of ${\rm Im}(\omega/\kappa_1)$  
resembles that of QNM 
frequencies of a Schwarzschild black hole. For the gravitational perturbations, 
the amplitude decreases for small values of 
${\rm Im}(\omega/\kappa_1)$, approaches the imaginary axis of the complex 
frequency plan, increases again, and finally approaches some constant value. 
The asymptotic value of the amplitude in the limit 
of highly damped modes seems to be non-zero constant, which is inferred as 
$\sim 0.4$. Therefore the limit of ${\rm Re}(\omega/\kappa_1)$ as 
${\rm Im}(\omega/\kappa_1)\rightarrow\infty$ appears not to exist for the gravitational 
perturbations. For the electromagnetic perturbations, on the other hand, the 
amplitude of oscillating ${\rm Re}(\omega/\kappa_1)$ decreases monotonically as 
${\rm Im}(\omega/\kappa_1)$ is increased. Its asymptotic value in the limit of 
highly damped modes seems to be zero. This means that for the electromagnetic 
perturbations, the limit of ${\rm Re}(\omega/\kappa_1)$ as ${\rm Im}(\omega/\kappa_1)$ 
goes to infinity seems to exist and to be zero. It is important to note that in a 
nearly extremal SdS black hole, the asymptotic behaviors of the quasinormal 
frequencies in the 
limit of highly damped modes are independent of the angular quantum number 
$l$ of the perturbations. Although we do not show the results for other 
values of $r_1\kappa_1$ and $l$, the asymptotic behavior of the QNM frequencies 
is not highly dependent on these parameters. 
In summary, our numerical results suggest that for the gravitational 
perturbations, the real parts of the QNM frequencies of nearly extremal 
SdS spacetimes do not go to any constant value in the limit of 
large imaginary frequencies because they show oscillating behaviors in 
the limit. For the electromagnetic perturbations, on the other hand, 
the real parts of the QNM frequencies seem to go to zero in the limit 
of large imaginary frequencies.

\section{Conclusions}
We have calculated the high-order QNMs with large imaginary frequencies 
for the electromagnetic and gravitational perturbations in nearly
extremal SdS spacetimes using  Leaver's continued 
fraction method \cite{leaver}. 
Our results show that for low-order QNMs, analytical formulas in 
the extremal limit derived by Cardoso and Lemos \cite{cardoso03} 
and Maassen van den Brink \cite{brink} is a quite good approximation 
for the QNM frequencies as long as the model parameter $r_1\kappa_1$ 
is small enough. 
For high-order QNMs, whose imaginary frequencies are 
sufficiently large, on the other hand, this formula becomes inaccurate 
even for small values of $r_1\kappa_1$. 
Therefore, the approximation derived by Cardoso and Lemos cannot give 
correct asymptotic behaviors of the QNMs in the limit of large
imaginary frequencies (see, also \cite{brink}). 
We also found that the real parts of the quasinormal frequencies 
have oscillating behaviors in the limit of highly damped modes.  
(Similar behaviors have been found in the quasinormal frequencies in a
Reissner-Nordstr{\" o}m black hole \cite{motl03,berti,an03,andersson_howls}.) 
The amplitude of oscillating ${\rm Re(\omega)}$ approaches a non-zero 
constant value for the gravitational perturbations and zero for the 
electromagnetic perturbations in the limit of highly damped modes, 
regardless of values of $l$ and $r_1\kappa_1$. This means that for the 
gravitational perturbations, the real parts of the quasinormal frequencies 
of nearly extremal SdS spacetimes appear not to go to any constant value 
in the limit of highly damped modes. Therefore our numerical results suggest 
that Hod's conjecture is not applicable for nearly extremal SdS black holes 
because the the limit of ${\rm Re(\omega)}$ as 
${\rm Im}(\omega)\rightarrow\infty$ does not exist.

Although we computed high-order QNMs whose damping rates are quite large, all the QNMs 
we obtained in this study are still associated with a finite mode number but not infinity, 
because we investigated the properties of the QNMs with straightforward numerical approach. 
Thus, we cannot exclude the possibility that our numerical results do not show correct 
asymptotic behaviors. 
Other approaches to examine asymptotic behaviors of QNMs in the highly damping 
limit are necessary, in order to confirm our results of the asymptotic behaviors. 
As for high-order QNMs with large imaginary frequencies for moderate values of 
$r_1\kappa_1$, Leaver's method cannot be applied straightforwardly. 
Therefore, other numerical techniques are needed to obtain QNMs with large imaginary 
frequencies.

\begin{acknowledgments} 
The authors are grateful to V. Cardoso for fruitful discussions and a 
careful reading of the manuscript. 
They are also grateful to J. Bekenstein for correspondence regarding 
the early history of the black hole thermodynamics.  S.Y. 
acknowledges financial support from Funda\c c\~ao para a  Ci\^encia e a 
Tecnologia (FCT) through project SAPIENS 36280/99.
\end{acknowledgments}

\appendix* 

\section{Superpartner and iso-spectral relationship between axial and polar perturbations in 
SdS spacetime} 

Non-radial gravitational perturbations of a SdS spacetime obey a Schr\"odinger-type wave 
equation, given by  
\begin{equation}
{d^2\phi^{(\pm)}\over dr_*^2}+[\omega^2-V^{(\pm)}(r)]\phi^{(\pm)}=0\,, 
\label{a1}
\end{equation}
where $\phi^{(+)}$ and $V^{(+)}$ ($\phi^{(-)}$ and $V^{(-)}$) are the gauge invariant perturbation 
function and the effective potential for polar (axial) parity perturbations, respectively. 
The effective potentials are given by 
\begin{eqnarray}
V^{(+)}&=&{2\Delta\over r^5(cr+3M)^2}\nonumber\\
&\times&[9M^3+9M^2cr+3c^2Mr^2\nonumber\\
&&\ \ +c^2(c+1)r^3-3M^2\Lambda r^3]\,, 
\label{a2}
\end{eqnarray}
\begin{eqnarray}
V^{(-)}={2\Delta\over r^5}\,[(c+1)r-6M]\,,
\label{a3}
\end{eqnarray}
where $c=(l+2)(l-1)/2$ (for detailed derivations of the master equation (\ref{a1}), see, e.g., 
 \cite{mm90,cl01}). As shown first by Cardoso and Lemos \cite{cl01}, two potentials $V^{(+)}$ 
and $V^{(-)}$ are simply related through the relation, given by
\begin{eqnarray}
V^{(\pm)}=\pm\beta {df\over dr_*}+\beta^2 f^2+\kappa f\,,
\label{a4}
\end{eqnarray}
where $\beta=6M$, $\kappa=4c(c+1)$, and 
\begin{equation}
f={\Delta\over 2r^3(cr+3M)}\,. 
\label{a5}
\end{equation}
This relation between two potentials are called the superpartner relationship. 
By virtue of the superpartner relationship, the perturbation function $\phi^{(+)}$ 
($\phi^{(-)}$) can be written in term of $\phi^{(-)}$ ($\phi^{(+)}$) and its first 
derivative \cite{ch83}, given by
\begin{eqnarray}
(\kappa\pm 2i\omega\beta)\phi^{(\pm)}=(\kappa+2\beta^2f)\,\phi^{(\mp)}\pm2\beta\,
{d\phi^{(\mp)}\over dr_*}\,. 
\label{a6}
\end{eqnarray} 
It is worthwhile to note that the superpartner relationship (\ref{a4}) and (\ref{a5}) and 
the relations between two functions $\phi^{\pm}$ (\ref{a6}) in a SdS spacetime have the same 
functional form as those in a Schwarzschild spacetime except for the definition of the 
function $\Delta$ \cite{ch83}. Since $\Delta=\Lambda(r-r_1)(r_2-r)(r-r_3)r/3$, the 
asymptotic form of the function $f$ in the limit of $r_*\rightarrow\pm\infty$ is 
given by 
\begin{equation}
f\rightarrow\left\{
\begin{array}{ll}
{\Lambda(r_2-r_1)(r_1-r_3)\over 6r_1^2(cr_1+3M)}\,e^{2|\kappa_1|r_*}&
{\rm as}\ r_*\rightarrow -\infty\\
{\Lambda(r_2-r_1)(r_2-r_3)\over 6r_2^2(cr_2+3M)}\,e^{-2|\kappa_2|r_*}&
{\rm as}\ r_*\rightarrow\infty\,. 
\end{array}
\right.
\end{equation}
Then, it is easy to see that the function $f$ has three properties; i) smooth for 
$-\infty< r_*< \infty$, ii) $f$ and its derivatives of all orders vanish as 
$r_*\rightarrow\pm\infty$, iii) an integral $\int^\infty_{-\infty}fdr_*$ exists.  
If the function $f$ appearing in (\ref{a4}) satisfies three conditions above, as shown in 
\cite{ch83}, two potentials $V^{(\pm)}$ give the same transmission amplitude and the 
same quasinormal frequencies. In a SdS spacetime, therefore, axial and polar perturbations 
yield the same set of quasinormal mode frequencies. This iso-spectral properties in a SdS 
spacetime is attributed to the fact that gravitational perturbations associated with 
a spin $s=-2$ in a SdS spacetime can be described with a single Weyl scalar $\Psi_4$ 
\cite{of91}.

In a Schwarzschild-anti-de Sitter spacetime, exactly the same relations between polar 
and axial perturbations (\ref{a4})--(\ref{a6}) obviously hold \cite{cl01}. Yet, 
there is no iso-spectral property between polar and axial perturbations in a 
Schwarzschild-anti-de Sitter spacetime. In a Schwarzschild-anti-de Sitter spacetime, 
the master equation (\ref{a1}) does not 
have an asymptotic solution given by $e^{\pm i\omega r_*}$ as $r\rightarrow\infty$ 
and, furthermore, $r_*$ has a finite range. The boundary condition at spatial 
infinity therefore must be modified.  One of the plausible boundary conditions is 
that perturbation functions vanish at spatial infinity, even though there are other 
options for the boundary conditions \cite{cl01}. If this boundary condition is taken 
at spatial infinity, in general, the transformation (\ref{a6}) between polar and 
axial perturbations cannot hold this boundary condition. Therefore, the set of the 
quasinormal frequencies of polar perturbations in a Schwarzschild-anti-de Sitter 
spacetime is not the same as that of axial perturbations.



\begin{figure}[ht]
\centering
\includegraphics[height=5cm,clip]{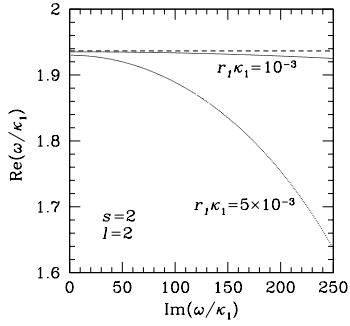}
\caption{Real parts of the non-dimensional QNM frequencies, $\omega/\kappa_1$, 
given as a function of the imaginary parts of the frequencies for the l=2 
gravitational perturbations. The model parameters $r_1\kappa_1$ are taken to be
$r_1\kappa_1=10^{-3}$ and $r_1\kappa_1=5\times 10^{-3}$. The frequencies 
obtained with the approximation formula in the limit of $r_1\kappa_1\rightarrow 0$ 
are also shown as the dashed curve.}
\end{figure}
\begin{figure}[ht]
\centering
\includegraphics[height=5cm,clip]{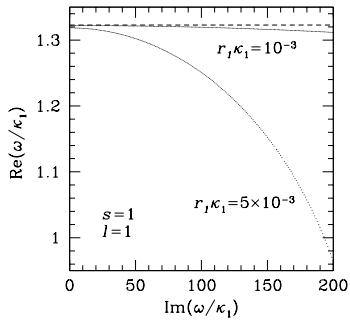}
\caption{Same as Figure 1 but for the $l=1$ electromagnetic perturbations.}
\end{figure}
\begin{figure}[ht]
\centering
\includegraphics[height=5cm,clip]{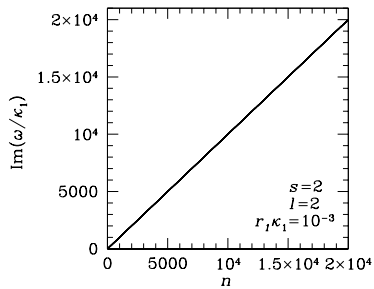}
\caption{Imaginary parts of the non-dimensional QNM frequencies, $\omega/\kappa_1$, 
given as a function of the mode number, $n$, for the gravitational perturbations 
associated with $l=2$. The model parameter $r_1\kappa_1$ is taken to be 
$r_1\kappa_1=10^{-3}$.}  
\end{figure}
\begin{figure}
\centering
\includegraphics[width=10cm]{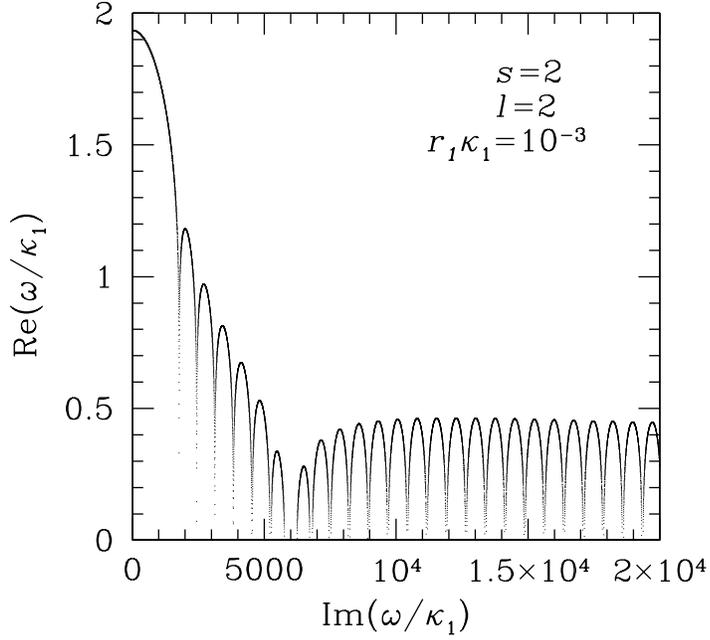}
\includegraphics[width=10cm]{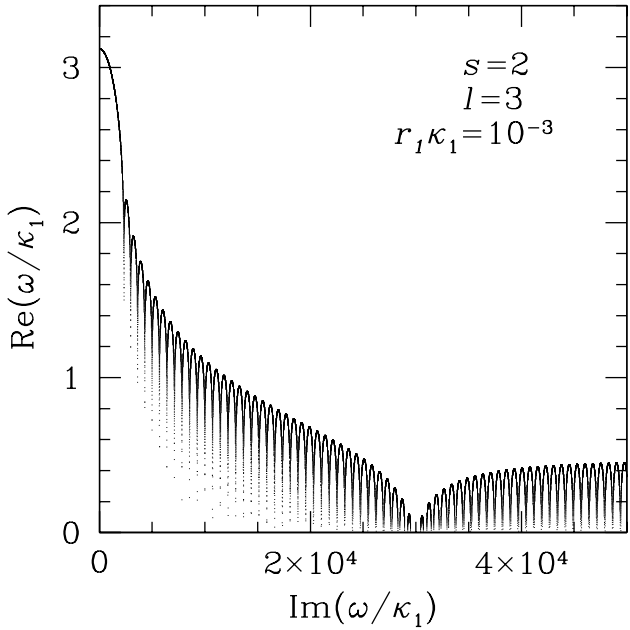}
\caption{Real parts of the non-dimensional QNM frequencies, $\omega/\kappa_1$,
given as a function of the imaginary parts of the frequencies for the 
gravitational perturbations having $l=2$ and $l=3$. The frequencies of the QNMs 
associated with different $l$, $l=2,\ 3$, are shown in each panel. The model 
parameters $r_1\kappa_1$ is taken to be $r_1\kappa_1=10^{-3}$.}
\end{figure}
\begin{figure}
\centering
\includegraphics[width=10cm]{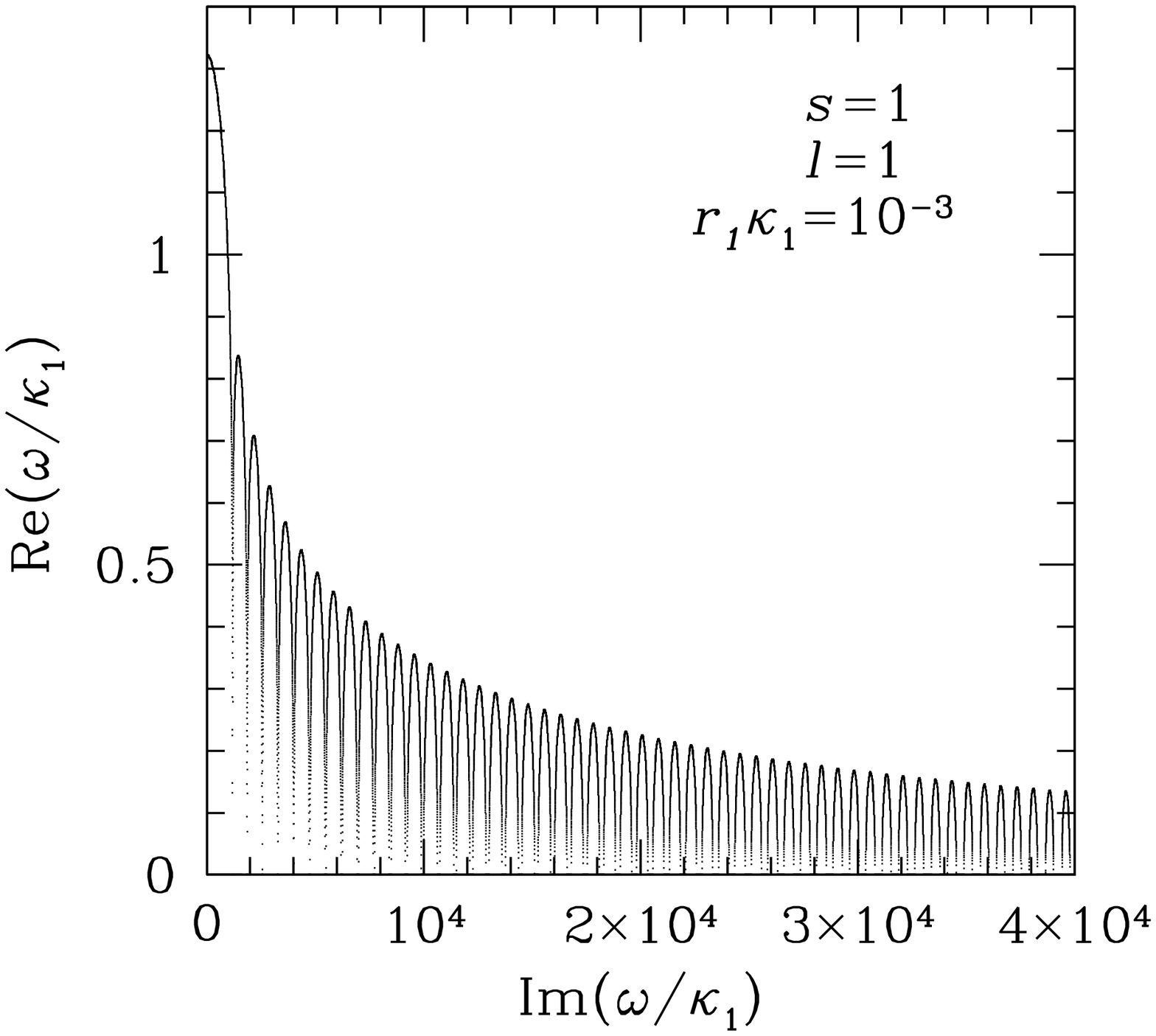}
\includegraphics[width=10cm]{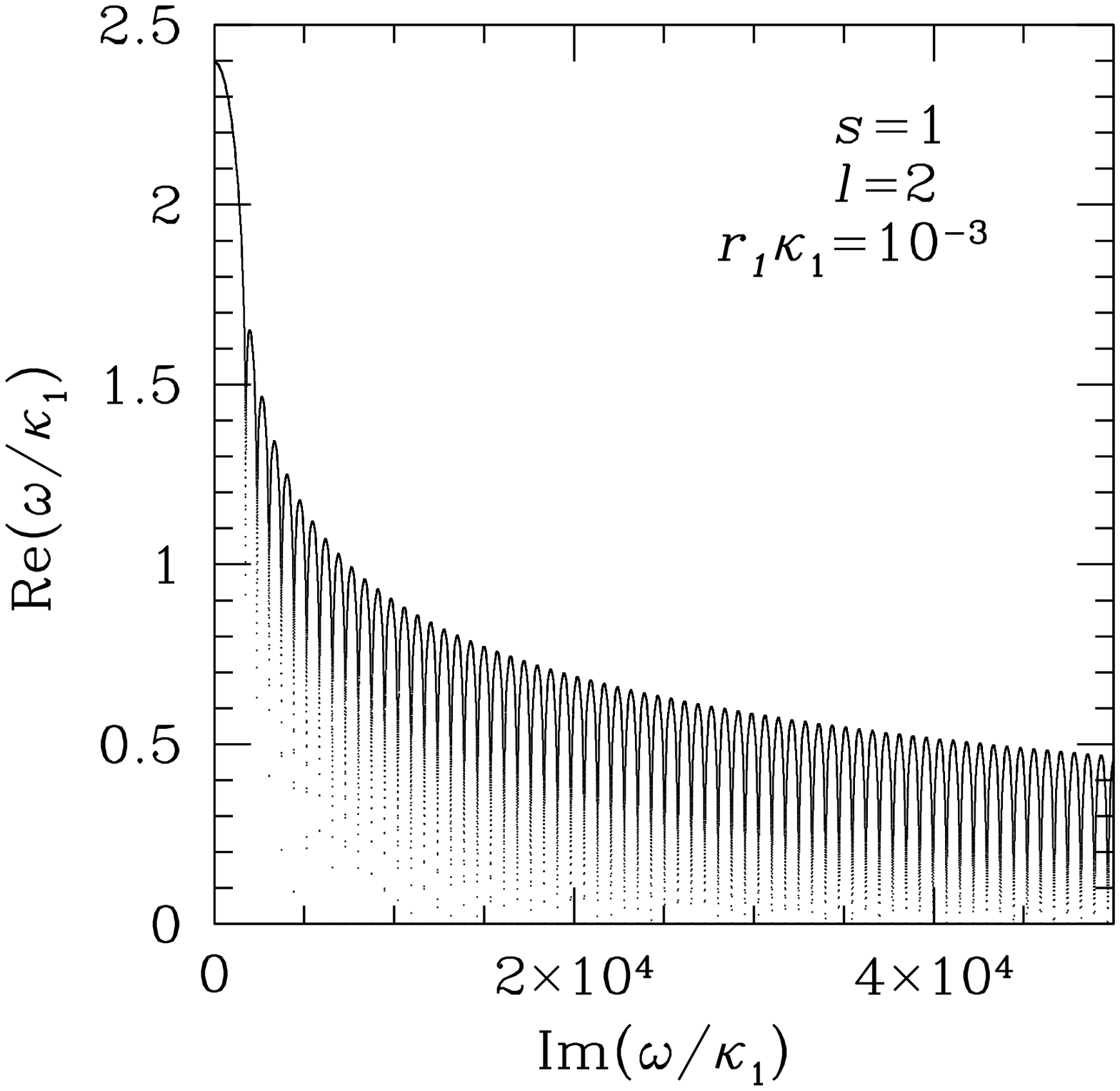}
\caption{Same as Figure 3 but for the electromagnetic perturbations having 
$l=1$ and $l=2$.}  
\end{figure}


\begin{thebibliography}{99}
%
\bibitem{vish} C. V. Vishveshwara, Nature {\bf 227}, 936 (1970). 
%
\bibitem{kokkotas99} K. D. Kokkotas and B. G. Schmidt, 
Living Rev. Relativity {\bf 2}, 2 (1999).
%
\bibitem{nollert99} H. P. Nollert, 
Class. Quant. Grav. {\bf 16}, R159 (1999).
%
\bibitem{MB95} J. D. Bekenstein and V. F. Mukhanov, Phys. Lett {\bf B 360},
 7 (1995). 
%
\bibitem{B97} J. D. Bekenstein, in {\it Proceedings of the Eight Marcel 
Grossmann  Meeting on General Relativity}, eds. T. Piran and R. Ruffini 
(World Scientific, Singapore 1999), pp. 92-111.
%
\bibitem{hod} S. Hod, Phys. Rev. Lett. {\bf 81}, 4293 (1998).
%
\bibitem{dreyer} O. Dreyer, Phys. Rev. Lett. {\bf 90}, 081301 (2003).
%
\bibitem{motl} L. Motl, Adv. Theor. Math. Phys. {\bf 6}, 1135 (2003).
%
\bibitem{nollert} H.-P. Nollert, Phys. Rev. D {\bf 47}, 5253 (1993).
%
\bibitem{motl03} L. Motl and A. Neitzke, Adv. Theor. Math. Phys. {\bf 7}, 
307 (2003). 
%
\bibitem{berti} E. Berti and K. D. Kokkotas, Phys. Rev. {\bf D} 68, 
044027 (2003).
%
\bibitem{an03} A. Neitzke, preprint (hep-th/0304080)
%
\bibitem{andersson_howls} N. Andersson and C. J. Howls, 
preprint (gr-qc/0307020). 
%
\bibitem{bertietal}E. Berti, V. Cardoso, K. D. Kokkotas, and 
H. Onozawa, Phys. Rev. D {\bf 68}, 124018 (2003).
%
\bibitem{hod03} S. Hod, Phys. Rev. D {\bf 67}, 081501 (2003). 
%
\bibitem{cardoso03} V. Cardoso and J. P. S. Lemos, Phys. Rev. D {\bf 67}, 
084020 (2003).
%
\bibitem{brink} A. Maassen van den Brink, Phys. Rev. D {\bf 68}, 
047501 (2003).
%
\bibitem{K02} R. A. Konoplya, Phys. Rev. D {\bf 66}, 044009 (2002).
%
\bibitem{CKL03} V. Cardoso, R. A. Konoplya, and J. P. S. Lemos,
Phys. Rev. D {\bf 68}, 044024 (2003).
%
\bibitem{leaver} E. W. Leaver,
Proc. R. Soc. London {\bf A402}, 285 (1985).
%
\bibitem{moss} I. G. Moss and J. P. Norman,
Class. Quant. Grav. {\bf 19}, 2323 (2002).
%
\bibitem{mm90}
F. Mellor and I. Moss, Phys. Rev. D {\bf 41}, 403 (1990); 
%
\bibitem{of91}
H. Otsuki and T. Futamase, Prog. Theor. Phys. {\bf 85}, 771 (1991); 
%
\bibitem{cl01}
V. Cardoso and J. P. S. Lemos, Phys. Rev. D {\bf 64}, 084017 (2001). 
%
\bibitem{gautschi} W. Gautschi, SIAM Rev. {\bf 9}, 24 (1967). 
%
\bibitem{andersson} N. Andersson, 
Proc. R. Soc. London {\bf A439}, 47 (1992).
%
\bibitem{cly03} V. Cardoso, J. P. S. Lemos, and S. Yoshida, Phys. 
Rev. D, in press (gr-qc/0309112); V. Cardoso, J. P. S. Lemos, and S. 
Yoshida, JHEP {\bf 0312}, 041 (2003). 
%
\bibitem{db03} D. Birmingham, Phys. Lett. {\bf B 569}, 199 (2003). 
%
\bibitem{ch83}
S. Chandrasekhar, {\it The Mathematical Theory of Black holes} 
(Oxford University, New York, 1983). 
\end{thebibliography}
\end{document}